\documentclass[aps,pra,twocolumn,superscriptaddress,showpacs,10pt]{revtex4-1}
\usepackage{amsmath,amssymb,graphicx,color}
\usepackage{epstopdf}
\begin{document}

\title{Coherent competition and control between three-wave mixing and four-wave mixing in superconducting circuits}

\author{Miao-Xiang Liang}
\affiliation{College of Physics and Electronic Science, Hubei Normal University, Huangshi 435002, China}

\author{Yu-Xiang Qiu}
\affiliation{College of Physics and Electronic Science, Hubei Normal University, Huangshi 435002, China}

\author{Hai-Chao Li}
\altaffiliation{hcl2007@foxmail.com}
\affiliation{College of Physics and Electronic Science, Hubei Normal University, Huangshi 435002, China}

\author{Wei Xiong}
\altaffiliation{xiongweiphys@wzu.edu.cn}
\affiliation{Department of Physics, Wenzhou University, Wenzhou 325035, China}

\begin{abstract}
Exploring intermixing and interplay between different frequency-mixing processes has always been one of the interesting subjects at the interface of nonlinear optics with quantum optics. Here we investigate coherent competition and control between three-wave mixing (TWM) and four-wave mixing (FWM) in a cyclic three-level superconducting quantum system. In the weak control-field regime, strong competition leads to an alternating oscillation between TWM and FWM signals and this oscillation is a signature of strong energy exchange between these two nonlinear processes. In particular, such oscillation is absent from conventional multi-wave mixing in atomic systems. Surprisingly, synchronous TWM and FWM processes are demonstrated in the strong control-field regime and, at the same time, their efficiencies can be as high as 40\% and 45\%, respectively. Our study shows that these competitive behaviors between TWM and FWM can be manipulated by tuning the control-field intensity.
\end{abstract}

\maketitle
Light-wave generation and control based on coherent atom-field interaction at the quantum level play a fundamentally important role in cavity quantum electrodynamics (QED)~\cite{Raimond,Walther,Reiserer}. In particular, nonlinear frequency mixing is a practical technique to produce and manipulate light signals~\cite{Fleischhauer}. Typically, four-wave mixing (FWM) with third-order nonlinearity has received a great deal of attention in atomic systems~\cite{Deng,Glasser,Li}. As examples, a large enhancement of FWM efficiency was reported in a $\Lambda$-type configuration~\cite{Liy}, a FWM process using electromagnetically induced transparency was demonstrated in cold atoms~\cite{Braje}, and an experimental observation of anti-parity-time symmetric FWM was performed in a four-level system~\cite{Jiang}. Moreover, six-wave mixing (SWM) was investigated in low-light-intensity limit~\cite{Kang} and microwave-to-optical conversion~\cite{Han}. Apart from individual FWM or SWM, coherent coexistence and spatiotemporal interference between FWM and SWM processes were experimentally realized in four-level atomic systems~\cite{Zhang,Zhang1,Zhang2}. However, all those works focus on 2n-wave mixing itself and intermixing between them. As a result, there remains a broad stage for the study of second-order nonlinear three-wave mixing (TWM) and its interaction with FWM. Unfortunately, exploiting the second-order nonlinearity in cavity QED encounters a long-standing challenge due to the presence of selection rules in natural atoms.

In parallel to atomic cavity QED, coherent interaction between superconducting artificial atoms~\cite{Clarke} and microwave photons allows to study quantum optics and quantum information in superconducting quantum circuits, which has led to the creation of the subject of circuit QED~\cite{You,Gu,Blais,Xue}. As artificially designed and fabricated quantum mechanical devices, circuit QED architectures open up new possibilities for tailoring and controlling fundamental light-matter interaction in unprecedented detail. For instance, a series of experiments in circuit QED showed the realization of strong, ultrastrong and deep-strong coupling regimes~\cite{Wallraff,Niemczyk,Yoshihara}. In addition, superconducting artificial atoms can be expected to possess new properties by suitably controlling circuit parameters. Beyond traditional electric-dipole selection rules, superconducting flux and fluxonium atoms~\cite{Liu,Manucharyan,Manucharyan1} can have a cyclic $\triangle$-type structure which does not exist in natural atoms. In terms of those outstanding advantages over cavity QED, circuit QED has been used to display strong nonlinear or less familiar effects, such as giant Kerr nonlinearities~\cite{Rebi}, multi-photon processes~\cite{Liu1}, and spontaneous virtual-photon emission~\cite{Stassi}. Especially, circuit QED with $\triangle$-type configuration offers an opportunity to study coexisting nonlinear wave-mixing processes including TWM.

Here we explore novel interplay between TWM and FWM processes in a cyclic $\triangle$-type superconducting quantum system. Several key features are emphasized in this work. First, simultaneous existence of TWM and FWM can result in coherent competition between these two nonlinear effects. Such competition has never been studied in atomic frequency mixing owing to the absence of TWM. Second, strong alternating oscillation between TWM and FWM signals emerges in the weak control-field condition, which is deemed to be a signature of efficient energy transfer between them. Such oscillation has not been reported in the coexistence of FWM and SWM in natural atoms~\cite{Zhang,Zhang1,Zhang2}. Third, in contrast with the weak-field case, nonlinear TWM and FWM processes present nearly synchronous evolution in the strong control-field regime. Meanwhile, the conversion efficiencies of both TWM and FWM can surprisingly reach $40\%$ and $45\%$, which are much larger than those of coexisting FWM and SWM (reported to be about $10\%$ and $1\%$) in atomic systems~\cite{Zhang}. Fourth, for these competitive TWM and FWM processes, a switch from the alternating oscillation to the synchronous evolution can be realized by changing the control field. Understanding and controlling these high-order frequency-mixing processes are of fundamental importance in the study of nonlinear optics and may have potential applications in many areas of physics.

\begin{figure}[t]
\centering{
\includegraphics[width=0.65\columnwidth]{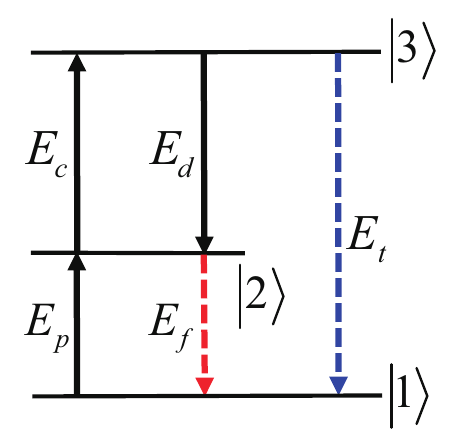}}
\caption{Schematic of a cyclic three-level superconducting quantum system driven by three incoming fields (solid lines). In this engineered light-matter interaction, two new signals $E_{t}$ and $E_{f}$ (dashed lines) can be generated by two coexisting nonlinear TWM and FWM processes, respectively.
}
\label{fig.1}
\end{figure}

Let us consider a cyclic three-level superconducting artificial system consisting of one ground state $|1\rangle$ and two excited states $|2\rangle$ and $|3\rangle$ with frequencies $\omega_{j}$ ($j=1,2,3$), as seen in Fig.~\ref{fig.1}. In the engineered light-matter interaction, a weak probe field $E_{p}$ (frequency $\omega_{p}$ and Rabi frequency $\Omega_{p}$) couples the transition $|1\rangle$ $\leftrightarrow$ $|2\rangle$ and a control field $E_{c}$ (frequency $\omega_{c}$ and Rabi frequency $\Omega_{c}$), together with a weak driving field $E_{d}$ (frequency $\omega_{d}$ and Rabi frequency $\Omega_{d}$), is applied to the transition $|2\rangle$ $\leftrightarrow$ $|3\rangle$. In this case, two coexisting nonlinear wave-mixing channels are opened by two different transitions $|1\rangle\rightarrow|2\rangle\rightarrow|3\rangle\rightarrow|1\rangle$ and $|1\rangle\rightarrow|2\rangle\rightarrow|3\rangle\rightarrow|2\rangle\rightarrow|1\rangle$. To be specific, a TWM process with second-order nonlinearity generates a new signal $E_{t}$ at a frequency $\omega_{t}=\omega_{p}+\omega_{c}$ and it can be described by a Liouville pathway $\rho^{(0)}_{11}\rightarrow\rho^{(1)}_{21}\rightarrow\rho^{(2)}_{31}$. Meanwhile, a FWM process with third-order nonlinearity generates a new signal $E_{f}$ at a frequency $\omega_{f}=\omega_{p}+\omega_{c}-\omega_{d}$ and it can be described by a Liouville pathway $\rho^{(0)}_{11}\rightarrow\rho^{(1)}_{21}\rightarrow\rho^{(2)}_{31}\rightarrow\rho^{(3)}_{21}$. Note that the control-field strength can be tuned from weak to strong in our nonlinear wave-mixing scheme.

In the rotating-wave approximation, our model can be described by the Hamiltonian ($\hbar=1$)
\begin{align}\label{Hamil}
H&=\sum_{j=1}^{3}\omega_{j}\sigma_{jj}-\frac{1}{2}(\Omega_{p}\rm{e}^{-i\omega_{p}t}\sigma_{21}+\Omega_{c}\rm{e}^{-i\omega_{c}t}\sigma_{32}\nonumber \\
&\quad+\Omega_{d}\rm{e}^{-i\omega_{d}t}\sigma_{32}+\Omega_{f}\rm{e}^{-i\omega_{f}t}\sigma_{21}+\Omega_{t}\rm{e}^{-i\omega_{t}t}\sigma_{31}\nonumber \\
&\quad+{\rm{H.c.}}),
\end{align}
where $\sigma_{ij}=|i\rangle\langle j|$ is the projection or transition operator of the superconducting artificial system and $\Omega_{t}$ ($\Omega_{f}$) is the Rabi frequency of the generated TWM (FWM) signal. Switching to the interaction picture, the Hamiltonian takes the form
\begin{align}\label{Hami2}
H_{I}&=\Delta_{p}\sigma_{22}+\Delta_{p}\sigma_{33}-\frac{1}{2}(\Omega_{p}\sigma_{21}+\Omega_{c}\sigma_{32}\nonumber \\
&\quad+\Omega_{d}\sigma_{32}+\Omega_{f}\sigma_{21}+\Omega_{t}\sigma_{31}+{\rm{H.c.}}),
\end{align}
where $\Delta_{p}=\omega_{21}-\omega_{p}$ is the detuning of the probe field and the frequency of the control (driving) field is assumed to match the energy spacing between the levels $|2\rangle$ and $|3\rangle$. After including the relaxation and dephasing processes, the superconducting atomic dynamics is governed by the Lindblad-type master equation
\begin{align}\label{three}
\frac{d\rho}{dt}&=-i[H_{I},\rho]+\frac{1}{2}\sum_{j=2}^{3}\gamma_{\phi j}(2\sigma_{jj}\rho\sigma_{jj}-\sigma_{jj}\rho-\rho\sigma_{jj})\nonumber \\
&\quad+\frac{1}{2}\sum_{i<j}\gamma_{ij}(2\sigma_{ij}\rho\sigma_{ji}-\sigma_{jj}\rho-\rho\sigma_{jj}),
\end{align}
where $\gamma_{\phi j}$ is the pure dephasing rate for the level $|j\rangle$ and $\gamma_{ij}$ is the relaxation rate between the levels $|i\rangle$ and $|j\rangle$. For a superconducting fluxonium system, $\gamma_{\phi j}$ is negligible in a wide range of flux around a degeneracy point~\cite{Manucharyan,Manucharyan1}. In a quantum mechanical treatment, linear and nonlinear atomic polarizations can be characterized by off-diagonal density matrix elements. According to Eqs.~(\ref{Hami2}) and~(\ref{three}), we have
\begin{align}\label{four}
{\dot{\rho}}_{21}=&\frac{i}{2}(\Omega_{p}+\Omega_{f})(\rho_{11}-\rho_{22})+\frac{i}{2}(\Omega^{*}_{c}+\Omega^{*}_{d})\rho_{31}-\frac{i}{2}\Omega_{t}\rho_{23}\nonumber \\
&-(\tau_{21}+i\Delta_{p})\rho_{21},\nonumber \\
{\dot{\rho}}_{31}=&\frac{i}{2}\Omega_{t}(\rho_{11}-\rho_{33})+\frac{i}{2}(\Omega_{c}+\Omega_{d})\rho_{21}-\frac{i}{2}(\Omega_{p}+\Omega_{f})\rho_{32}\nonumber \\
&-(\tau_{31}+i\Delta_{p})\rho_{31},\nonumber \\
{\dot{\rho}}_{32}=&\frac{i}{2}(\Omega_{c}+\Omega_{d})(\rho_{22}-\rho_{33})-\frac{i}{2}(\Omega^{*}_{p}+\Omega^{*}_{f})\rho_{31}+\frac{i}{2}\Omega_{t}\rho_{12}\nonumber \\
&-\tau_{32}\rho_{32},
\end{align}
where $\tau_{21}=\frac{1}{2}(\gamma_{12}+\gamma_{\phi2})$, $\tau_{31}=\frac{1}{2}(\gamma_{13}+\gamma_{23}+\gamma_{\phi3})$, and $\tau_{32}=\frac{1}{2}(\gamma_{12}+\gamma_{13}+\gamma_{23}+\gamma_{\phi2}+\gamma_{\phi3})$. Assuming that the system is initially prepared in the ground state $|1\rangle$, the steady-state off-diagonal elements associated with the two transition channels $|1\rangle$ $\leftrightarrow$ $|2\rangle$ and $|1\rangle$ $\leftrightarrow$ $|3\rangle$ are given by a formal perturbation expansion
\begin{subequations}\label{five}
\begin{align}
\rho^{(1)}_{31}&=\frac{i\Omega_{t}\Gamma_{21}}{2\lambda},\\
\rho^{(1)}_{21}&=\frac{i(\Omega_{f}+\Omega_{p})\Gamma_{31}}{2\lambda},\\
\rho^{(2)}_{31}&=\frac{i^{2}\Omega_{p}\Omega_{c}}{4\lambda}+\frac{i^{2}\Omega_{f}\Omega_{d}}{4\lambda},\\
\rho^{(2)}_{21}&=\frac{i^{2}\Omega_{t}\Omega^{*}_{c}}{4\lambda}+\frac{i^{2}\Omega_{t}\Omega^{*}_{d}}{4\lambda},\\
\rho^{(3)}_{21}&=\frac{i^{3}\Omega_{p}\Omega_{c}\Omega^{*}_{d}}{8\lambda\Gamma_{21}}+\frac{i^{3}\Omega_{f}\Omega_{d}\Omega^{*}_{c}}{8\lambda\Gamma_{21}},
\end{align}
\end{subequations}
where $\Gamma_{21}=\tau_{21}+i\Delta_{p}$, $\Gamma_{31}=\tau_{31}+i\Delta_{p}$, and $\lambda=\Gamma_{21}\Gamma_{31}+|\Omega_{c}|^2/4$. Equations~(\ref{five}a) and~(\ref{five}b) are responsible for the linear absorptions of the TWM, FWM and probe fields, respectively. The first terms in Eqs.~(\ref{five}c) and~(\ref{five}d) represent the nonlinear TWM and its backward nonlinear process via reabsorption, while the second terms illustrate the mutual frequency conversions between the TWM and FWM signals. Similarly, the two terms in Eq.~(\ref{five}e) describe the nonlinear FWM and its backward nonlinear process. Using the slowly varying amplitude approximation~\cite{Shen}, a set of propagation equations for the probe field and the generated wave-mixing signals can be written as
\begin{subequations}\label{six}
\begin{align}
\frac{\partial\Omega_{p}}{\partial z}&=i\kappa_{12}\left[\frac{i\Gamma_{31}\Omega_{p}}{2\lambda}+\frac{i^{2}\Omega_{t}\Omega^{*}_{c}}{4\lambda}+\frac{i^{3}\Omega_{f}\Omega_{d}\Omega_{c}^{*}}{8\Gamma_{21}\lambda}\right], \\
\frac{\partial\Omega_{t}}{\partial z}&=i\kappa_{13}\left[\frac{i\Gamma_{21}\Omega_{t}}{2\lambda}+\frac{i^{2}\Omega_{p}\Omega_{c}}{4\lambda}+\frac{i^{2}\Omega_{f}\Omega_{d}}{4\lambda}\right], \\
\frac{\partial\Omega_{f}}{\partial z}&=i\kappa_{12}\left[\frac{i\Gamma_{31}\Omega_{f}}{2\lambda}+\frac{i^{2}\Omega_{t}\Omega^{*}_{d}}{4\lambda}+\frac{i^{3}\Omega_{p}\Omega_{c}\Omega^{*}_{d}}{8\Gamma_{21}\lambda}\right],
\end{align}
\end{subequations}
where $\kappa_{ij}$ is a propagation constant. Obviously, the coupled equations demonstrate that not only can the probe field generate the wave-mixing signals, but also the generated signals can have a significant influence on the probe field during propagation. In particular, the coexistence and intermixing of nonlinear TWM and FWM processes in the superconducting $\triangle$-type system would lead to novel competitive effects which can not occur in natural atoms due to the absence of TWM.

\begin{figure}[b]
\centerline{
\includegraphics[width=0.98\columnwidth]{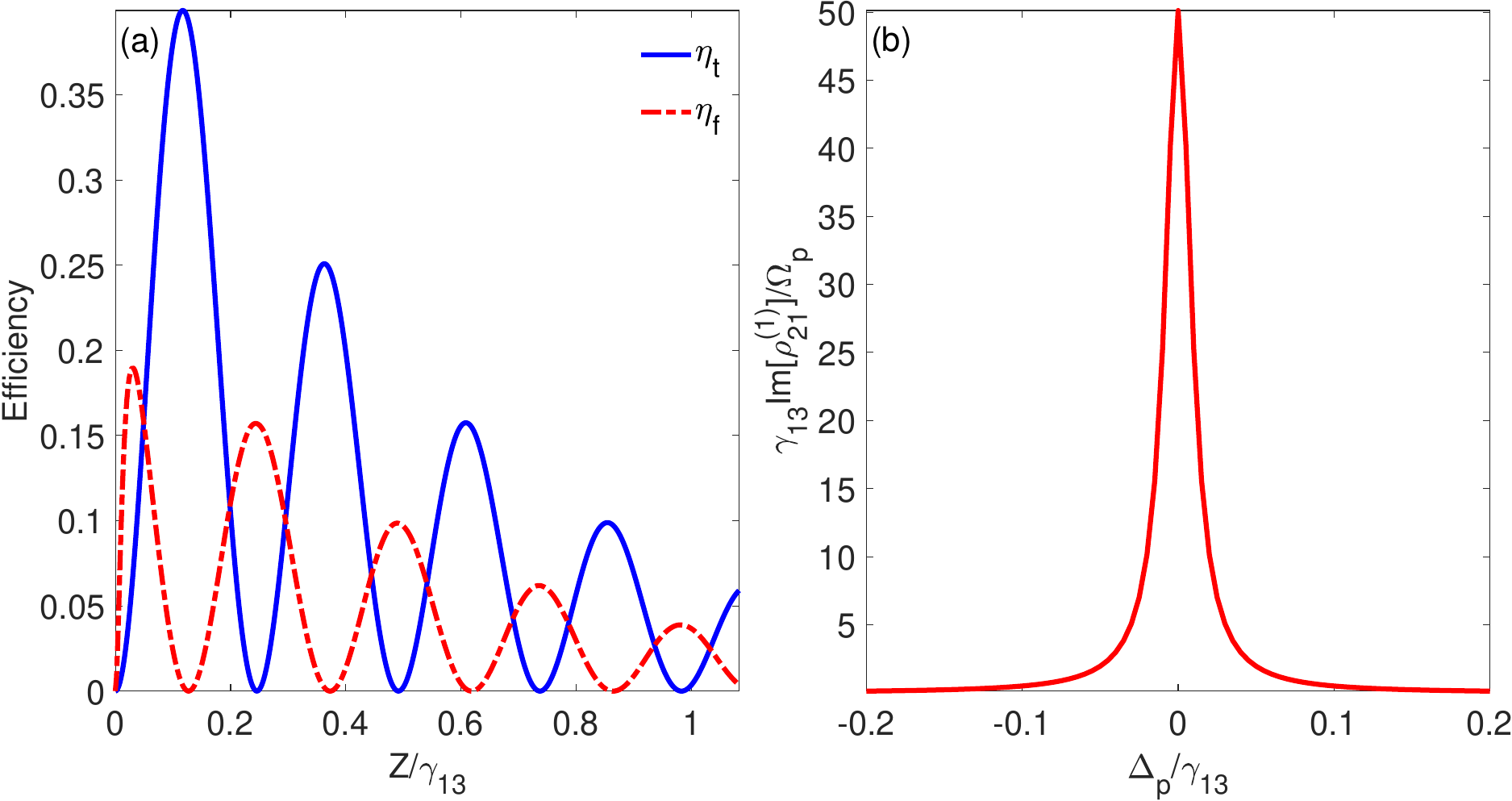}}
\caption{(a) TWM (blue solid line) and FWM (red dashed-dotted line) efficiencies as a function of the effective propagation distance $Z$ in the weak control-field regime. (b) Linear absorption of the probe field versus the probe-field detuning $\Delta_{p}$. The parameters are $\Omega_{c}=\Omega_{d}=0.1\gamma_{13}$, $\gamma_{12}=0.01\gamma_{13}$, $\gamma_{23}=0.005\gamma_{13}$, $\kappa_{13}=3.3\kappa_{12}$, $\mu_{13}=\mu_{12}$, $\Delta_{p}=0$, and $\omega_{t}=3.3\omega_{p}$.
}
\label{fig.2}
\end{figure}

\begin{figure}[b]
\centerline{
\includegraphics[width=0.981\columnwidth]{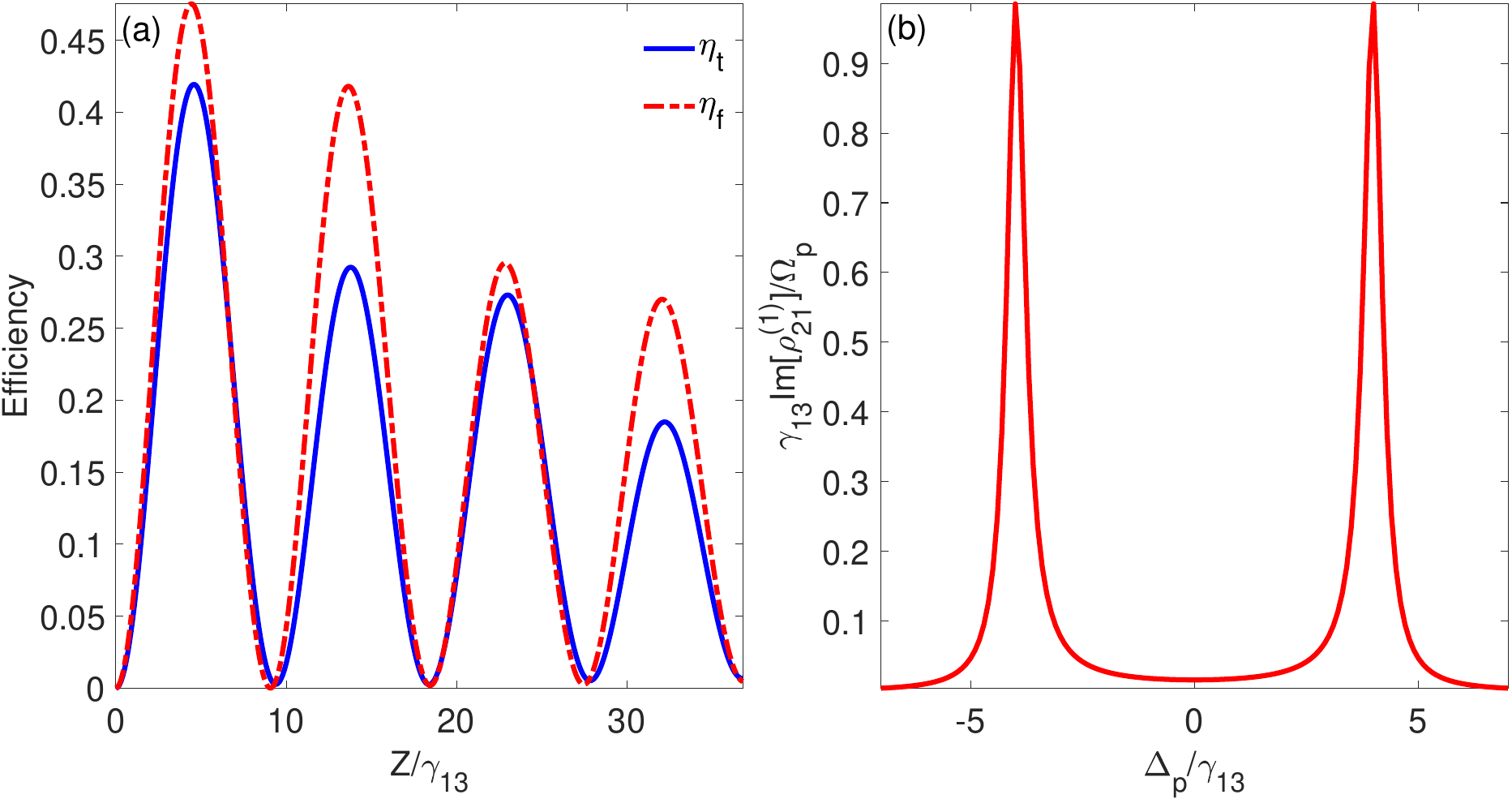}}
\caption{(a) TWM (blue solid line) and FWM (red dashed-dotted line) efficiencies as a function of the effective propagation distance $Z$ in the strong control-field regime. (b) Linear absorption of the probe field versus the probe-field detuning $\Delta_{p}$. The parameters are $\Omega_{c}=8\gamma_{13}$, $\Omega_{d}=0.65\gamma_{13}$, $\gamma_{12}=0.01\gamma_{13}$, $\gamma_{23}=0.005\gamma_{13}$, , $\kappa_{13}=3.3\kappa_{12}$, $\mu_{13}=\mu_{12}$, $\Delta_{p}=0.16\gamma_{13}$, and $\omega_{t}=3.3\omega_{p}$.}
\label{fig.3}
\end{figure}

Equations~(\ref{six}a)--(\ref{six}c) can be solved exactly by a standard procedure in differential-equation theory. However, the analytical solutions are rather complicated and tedious, which makes them have no significance in physics. So it is not necessary to give these analytical expressions in this work. Alternatively, we present numerical results for these equations. Figure~\ref{fig.2}(a) plots the conversion efficiency of nonlinear frequency mixing versus the effective propagation distance $Z=\kappa_{12}z$ in the weak control-field condition $\Omega_{c}=0.1\gamma_{13}$. Note that the efficiency is defined as the photon-number ratio of the generated wave-mixing signal to the incident probe field~\cite{Harris}. According to this definition, the TWM and FWM efficiencies are given by $\eta_{t}=|\mu_{12}\Omega_{t}/(\mu_{13}\Omega_{p0})|^{2}\omega_p/\omega_t$ and $\eta_{f}=|\Omega_{f}/\Omega_{p0}|^{2}$, where $\Omega_{p0}$ is the initial Rabi frequency of the probe field at $z=0$ and $\mu_{12}$ ($\mu_{13}$) is the dipole moment of the transition $|1\rangle\leftrightarrow|2\rangle$ ($|1\rangle\leftrightarrow|3\rangle$). This figure shows a strong alternating oscillation between TWM and FWM efficiencies. Specifically, when the TWM efficiency reaches its maximum value, the FWM efficiency takes its minimum value, and vice versa. In physics, such oscillation is a signature of efficient energy transfer between TWM and FWM processes and this is totally different from the coexisting FWM and SWM situation~\cite{Zhang1} in atoms, where FWM and SWM signals reach their respective steady-state values during propagation. Here the paired maximum-minimum values in the efficiency curves can be explained as a result of severe loss of the weak probe field. Obviously, the probe field undergoes significant linear absorption in Fig.~\ref{fig.2}(b) and thus can not provide sufficient photons to participate in the simultaneous TWM and FWM processes. When all probe photons are converted into the TWM (FWM) signal, the FWM (TWM) process vanishes, thereby giving rise to the paired maximum-minimum efficiencies in nonlinear frequency conversions.

On the contrary, these two nonlinear wave-mixing processes in Fig.~\ref{fig.3}(a) exhibit almost synchronous evolution in the strong-field regime $\Omega_{c}=8\gamma_{13}$, as the effective propagation distance $Z$ increases. The occurrence for the pairs of maximum-maximum values in the curves can be illustrated again by the linear absorption of the probe field. In Fig.~\ref{fig.3}(b), the probe field displays a clear transparency window in spectrum and its linear absorption can be greatly suppressed. In this case, there are a large number of probe photons to take part in nonlinear mixing processes, which opens the simultaneous TWM and FWM channels. Moreover, we should stress that one of the central topics in frequency mixing is how to obtain large conversion efficiency in multi-level quantum systems. Surprisingly, although two nonlinear TWM and FWM processes coexist in our scheme, their efficiencies can be as high as $40\%$ and $45\%$, respectively, as shown in Fig.~\ref{fig.3}(a). These efficiencies are much greater than those of coexisting FWM and SWM (about $10\%$ and $1\%$) in atomic systems~\cite{Zhang}. In addition, here the total conversion efficiency from the probe field to the wave-mixing signals reaches $85\%$, which is also remarkably larger than that of conventional frequency-mixing scheme in atoms~\cite{Li,Liy,Braje}.

In conclusion, we developed the theory of coexisting frequency-mixing processes at the quantum level by introducing second-order nonlinear TWM. We demonstrated the coherent interplay and control between TWM and FWM processes in a cyclic $\triangle$-type superconducting artificial system. Specifically, strong alternating oscillation and energy exchange between these two nonlinear processes are presented in the weak-field regime while synchronous evolution between them, as well as high efficiencies of $40\%$ and $45\%$, is realized in the strong-field region. These competitive behaviors can be manipulated by adjusting the control-field intensity and can be well explained by analyzing the linear absorption of the probe field. Our study is quite different from the coexisting FWM and SWM situation previously reported in natural atoms. Manipulating and optimizing these nonlinear TWM and FWM processes may have important applications in coherent quantum control, nonlinear optical spectroscopy, and quantum information science.

This work was partially supported by the Natural Science Foundation of Hubei Province of China under Grant No. 2022CFB509 and the National Natural Science Foundation of China under Grant No. 11904201.

\end{document}